# Quantifying the structural integrity of nanorod arrays

Florian Thöle,[§] Longjian Xue,[*] Claudia Heß,[§] Reinald Hillebrand,[$] Stanislav N. Gorb,[&] Martin Steinhart[§]

[§] Institut für Chemie neuer Materialien der Universität Osnabrück, Barbarastr. 7, 49069 Osnabrück, Germany

[*] School of Power and Mechanical Engineering; Wuhan University; South Donghu Road 8, Wuhan, Wuchang, 430072; Hubei, China

[&] Functional Morphology and Biomechanics, Zoological Institute, Kiel University, Am Botanischen Garten 9, 24118 Kiel, Germany

[$] Max-Planck-Institut für Mikrostrukturphysik, Weinberg 2, 06120 Halle, Germany

Correspondence to: Longjian Xue, School of Power and Mechanical Engineering; Wuhan University; Donghu South Road 8, Wuhan, Wuchang, 430072; Hubei, China, e-mail longjian.xue@uos.de; Martin Steinhart, Institut für Chemie neuer Materialien, Universität Osnabrück, Barbarastr. 7, 49069 Osnabrück, e-mail: martin.steinhart@uos.de

**Summary**

Arrays of aligned nanorods oriented perpendicular to a support, which are accessible by top-down lithography or by means of shape-defining hard templates, have received increasing interest as sensor components, components for nanophotonics and nanoelectronics, substrates for tissue engineering, surfaces having specific adhesive or antiadhesive properties and as surfaces with customized wettability. Agglomeration of the nanorods deteriorates the performance of components based on nanorod arrays. A comprehensive body of literature deals with mechanical failure mechanisms of nanorods and design criteria for mechanically stable nanorod arrays. However, the structural integrity of nanorod arrays is commonly evaluated only visually and qualitatively. We use real-space analysis of microscopic images to quantify the fraction of condensed nanorods in nanorod arrays. We suggest the number of array elements apparent in the micrographs divided by the number of array elements a defect-free array would contain in the same area, referred to as integrity fraction, as a measure of structural array integrity. Reproducible procedures to determine the imaged number of array elements are introduced. Thus, quantitative comparisons of different nanorod arrays, or of one nanorod array at different stages of its use, are possible. Structural integrities of identical nanorod arrays differing only in the length of the nanorods are exemplarily analyzed.

**Introduction**

Arrays of aligned nanorods oriented perpendicular to a substrate, which are accessible by top-down lithography (del Campo & Arzt, 2008; Pokroy et al., 2009a; Boesel et al., 2010), the use of shape-defining hard templates (Martin, 1994) or hot embossing (Jeong et al., 2006), have been used as surfaces with specific adhesive properties (Lee et al., 2007; Ho et al., 2011; Kwak et al., 2011; Xue et al., 2013; Xue et al., 2015) or specific wettability (Jin et al., 2005; Feng & Jiang, 2006), as substrates for tissue engineering (Kim et al., 2006; Steinberg et al., 2007; Saez et al., 2007; Mahdavi et al., 2008; Kim et al., 2009; Grimm et al., 2010), as drug delivery systems (Tao & Desai, 2007; Bernards & Desai, 2010), as templates in the generation of inorganic specimens with porous surfaces (Yanagishita et al., 2005; Grimm et al., 2008), as potential components for organic photovoltaics (Haberkorn et al., 2009) and as sensor components (Gitsas et al., 2010). Agglomeration of the nanorods deteriorates the performance of the nanorod arrays in the above-mentioned applications. For example, buckling of nanorods in adhesive systems results in reduced adhesion (Glassmaker et al., 2004). The mechanisms leading to the collapse of nanorods, predominantly the occurrence of capillary forces during evaporation of liquids covering the nanorod arrays (Tanaka et al., 1993; Chandra et al., 2009; Chandra & Yang, 2009) and nanorod/substrate adhesion as well as adhesion between adjacent nanorods, have been intensively studied (Chandra & Yang 2010). Various strategies for the design of mechanically stable nanorod arrays by tailoring aspect ratio (Roca-Cusachs et al.; 2005; Zhang et al., 2006; Choi et al., 2011), stiffness (Zhang et al., 2006) and surface chemistry (Matsunaga et al., 2011) of the nanorods were reported. Given the comprehensive body of literature dealing with the fabrication and the mechanical stability of nanorod arrays, it is astounding that their structural integrity has predominantly been evaluated in a qualitative way by visual inspection of microscopic images. However, the assessment of quality, performance and usability of nanorod arrays requires quantitative statistical analysis of their structural integrity, which obviously deteriorates if the nanorods agglomerate. Up to now, real-space image analysis has predominantly dealt with statistical grain analysis (Mátéfi-Tempfli et al., 2008; Hillebrand et al., 2008; Johnston-Peck et al., 2011; Beck & Bretzler 2011) and with the determination of the fraction of broken microrods in microrod arrays (Hillebrand et al., 2009). An elegant method for addressing array elements in arrays without long-range order, as typically obtained by bottom-up methods, was reported by Vlad et al. (Vlad et al., 2010). However, real space image analysis of corresponding micrographs as a reliable means to reproducibly quantify the structural integrity of nanorod arrays has remained unexplored.

Here, we suggest the number of array elements $N_{IF}$ apparent in microscopic images of nanorod arrays divided by the number $N_0$ of array elements ideal, defect-free counterparts of the pictured nanorod arrays would contain as a quantitative measure for the structural integrity of nanorod arrays, which is, in the following, referred to as integrity fraction $IF$. Thus, the structural integrities of either different nanorod arrays or of one specific nanorod array at different stages of its use can be quantitatively compared. As discussed below in detail, we developed procedures based on real-space image analysis of microscopic images of nanorod arrays for the reproducible determination of $N_{IF}$ and, therefore, $IF$.

**Model system**

As example in case, we evaluated the integrity of arrays of poly(methyl methacrylate) (PMMA) nanorods 180 nm in diameter imaged by scanning electron microscopy (SEM) (Seiler, 1983). SEM typically yields electronic grayscale images composed of a certain number of picture elements (pixels). Commonly, the image depth amounts to 8 bits so that one out of 256 brightness levels (pixel intensities) can be assigned to a pixel. Each pixel corresponds to one scan point, and its pixel intensity encodes the locally measured properties of the sample under investigation. Owing to several contrast mechanisms, in particular the edge effect, nanorod tips appear bright in secondary electron images.

The PMMA nanorod arrays were obtained by replicating self-ordered nanoporous alumina (anodic aluminum oxide, AAO) (Masuda et al., 1998) containing straight, aligned cylindrical nanopores with a di-ameter of ~180 nm. The AAO nanopore arrays with a nominal lattice constant of ~500 nm consisted of grains extending 10 to 20 lattice periods, in which the nanopores formed hexagonal lattices (Nielsch et al., 2002). While the nanorod arrays were otherwise prepared in the same way using the same materials (cf. Materials and Methods section), the depths of the AAO nanopores, and consequently the lengths $L$ of the PMMA nanorods, were varied from 1 µm (Figure 1A) to 1.5 µm (Figure 1B) to 2 µm (Figure 1C) to 5 µm (Figure 1D).

The set of SEM images seen in Figure 1 has exemplary properties suitable to demonstrate quantitative assessment of the structural integrities of the imaged nanorod arrays. The nanorod array with $L = 1$ µm seen in Fig. 1A predominantly consists of separate nanorods and is an intact negative replica of the pore array of the AAO template used in its preparation. Regarding the quality of both pictured sample and the image itself, Figure 1A is close to ideal. Figure 1B shows a largely intact nanorod array in which a few small clusters of condensed nanorods appear. A second important feature of Figure 1B is the high level of pixel noise (statistical brightness fluctuations on short length scales of a few pixels) in the background surrounding the nanorods. Figure 1C shows a nanorod array in which a significant fraction of nanorods is clustered, albeit small areas consisting of separate nanorods also exist. The fraction of clustered nanorods increases and the fraction of separated nanorods correspondingly decreases from Figure 1A to Figure 1B to Figure 1C. In Figure 1D virtually all nanorods are incorporated into large nanorod aggregates. While Figure 1A comes close to the limiting case of a nearly perfectly ordered nanorod array, Figure 1D corresponds to the limiting case of complete structural collapse.

For ideal nanorod arrays, in which the nanorods form defect-free hexagonal lattices with lattice constant $d$, the number of nanorods $N_0$ within an area $A$ can be calculated according to

$$N_0 = \frac{2 \cdot A}{d^2 \cdot \sqrt{3}}.$$

By application of the procedures described in the following, $N_{IF}$ and the *IF* values can be determined. Table 1 summarizes the results thus obtained for the PMMA nanorod arrays displayed in Figure 1. The $N_{IF}$ value obtained for Figure 1A is slightly larger than $N_0$ (and the *IF* value thus obtained is slightly larger than 1). This outcome might be related to an actual mean lattice constant of the imaged PMMA nanorod array that is slightly smaller than the assumed lattice constant of 500 nm.

**Thresholding**

Before objects in a microscopic image, such as single nanorods or aggregates of nanorods, can be counted, they need to be separated from the background by a thresholding step. The histogram of a digital image displays how many pixels have a specific pixel intensity. If a threshold value $I_T$ for the pixel intensity is set, all pixels having pixel intensities equal or larger than $I_T$ are considered white, while pixels having smaller pixel intensities than $I_T$ are considered black. Black areas are discarded; discrete areas consisting of contiguous white pixels represent countable objects. The question arises how the selection of $I_T$ influences the number of recognized objects $N_R$. As described previously (Hillebrand et al., 2009), $N_R(I_T)$ profiles are obtained as follows. $I_T$ is successively increased from 0 to 255. Areas having pixel intensities below $I_T$ are discarded, and the remaining discrete objects consisting of contiguous pixels having pixel intensities larger than $I_T$ are counted as a function of $I_T$.

Figure 2A shows an ideal hexagonal model array consisting of $N_R = N_0 = 2060$ array elements (corresponding to nanorod tips) with a pixel intensity of 198 surrounded by background with a pixel intensity of 49, and Figure 2B the corresponding $N_R(I_T)$ profile. Background and array elements are counted as one single entity as long as $I_T$ is smaller than or equal to 49. As soon as $I_T$ is equal or larger than 50, the background is discarded, and each of the brighter array elements now separated from the background is recognized as a single object. Thus, $N_R$ increases from one to 2060 in a discrete step. As soon as $I_T$ becomes larger than 198, the array elements merge into the background and $N_R$ decreases in another dis-crete step from $N_0$ to zero. Hence, for ideal arrays of elements uniform in brightness surrounded by a darker, likewise uniform background nearly rectangular $N_R(I_T)$ profiles with a plateau between the $I_T$ values at which background and array elements are discarded result.

**Object size**

The situation depicted in Figure 2 is idealized in that both background and array elements are uniform in pixel intensity. In real microscopic images, pixel noise superimposes on image features originating from sample topography. The consequences are exemplarily outlined in Figure 3, which shows a detail of Figure 1B at $I_T = 67$ (Figure 3A), $I_T = 95$ (Figure 3B) and $I_T = 133$ (Figure 3C). Figure 3A contains 264 objects consisting of one or several contiguous white pixels. However, whereas only 18 objects have areas of 25 contiguous white pixels and above, 164 objects consist of only one white pixel; the vast majority of identified objects originates from random local brightness fluctuations (pixel noise). 10 discrete objects representing single nanorod tips have areas ranging from than ~120 to ~190 white pixels. Some of the nanorod tips are connected by thin strings of white pixels that can be ascribed rather to background noise than to sample topography. Therefore, four large objects containing up to six nanorod tips (bottom right of Figure 3A) occur. The identified objects in the background resulting from pixel noise have lower pixel intensities than the pixels representing the nanorod tips. Hence, most of the small objects consisting of one white pixel or of a few contiguous white pixels vanish as $I_T$ is increased. Likewise, the thin strings between some of the nanorod tips disappear so that the nanorod tips become separated. In Figure 3B, only 58 discrete objects appear, 25 of which consist of a single white pixel. 21 objects with sizes ranging from 61 to 94 white pixels represent the nanorod tips completely or largely located in the image field, which are now all separated from each other. If $I_T$ is increased to 133, the larger entities representing the nanorod tips start to dissolve into smaller objects. Thus, Figure 3C contains 78 discrete objects, 34 of which consist of a single white pixel while not a single entity containing 25 or more contiguous white pixels exists. It is obvious that, besides $I_T$, the minimum number $S_M$ of contiguous white pixels considered and counted as an object strongly influences $N_R$. If, for example, $S_M$ is set to one, every single white pixel will be considered as an object and counted. However, if $S_M$ is set to 20, only objects consisting of at least 20 contiguous white pixels will be counted.

**$N_R(I_T,S_M)$ surfaces**

The properties of micrographs showing nanorod arrays are best represented by plots of $N_R$ as a function of $I_T$ and $S_M$ (Figure 4). All $N_R(I_T,S_M)$ surfaces have in common that $N_R$ steeply decreases if $S_M$ is increased departing from $S_M = 1$ as small objects consisting of one white pixel, or a few contiguous white pixels, are no longer counted. This region of the $N_R(I_T,S_M)$ surfaces is bordered by a sharp kink normal to the $S_M$ direction and roughly parallel to the $I_T$ direction. The slope of the $N_R(I_T,S_M)$ surfaces along the $S_M$ direction is much smaller at the larger $S_M$ values beyond the sharp kink than at the smaller $S_M$ values before the sharp kink. For $S_M$ values beyond the sharp kink, the $N_R(I_T)$ contour lines at fixed $S_M$ parallel to the kink are realistic versions of the ideal $N_R(I_T)$ profile shown in Figure 2B. $N_R(I_T)$ is small at small and large $I_T$ values, while $N_R(I_T)$ reaches a maximum at intermediate $I_T$ values.

The $N_R(I_T,S_M)$ surfaces belonging to Figure 1A shown in Figure 4A, to Figure 1B shown in Figure 4B and to Figure 1C shown in Figure 4C exhibit triangular plateaus representing stationary states in which $N_R$ remains constant or changes only slightly if $I_T$ and/or $S_M$ are varied. The plateau in Figure 4A has a height of $N_R \approx 2150$, a value in good agreement with $N_0 = 2056$. The plateau in Figure 4B has a lower height of $N_R \approx 1600$ and that in Figure 4C an even lower height of $N_R \approx 840$, while $N_0$ still amounts to 2056 in both cases. In the case of Figures 4B and 4C, the difference between $N_0$ and $N_R$ approximately represents the fraction of nanorods condensed to aggregates counted as single entities. The $N_R(I_T,S_M)$ surface belonging to Figure 1D (Figure 4D) shows qualitatively different features. At small $S_M$ values, $N_R$ steeply decreases as $S_M$ increases. At $S_M \approx 20$, the $N_R(I_T,S_M)$ surface has a sharp kink and passes into a broad, flat foot-like structure with corrugations roughly parallel to the $S_M$ axis. The height of the flat foot slowly decreases with increasing $S_M$ starting from a maximum value of $N_R \approx 160$.

**Slope magnitude maps of the $N_R(I_T,S_M)$ surfaces**

Figure 5 shows maps displaying the local magnitudes of the difference quotients of the functions $N_R(I_T,S_M)$, which represent, in good approximation, the local magnitude of slope of the $N_R(I_T,S_M)$ surfaces seen in Figure 4. The slope magnitude map (Figure 5A) of the $N_R(I_T,S_M)$ surface (Figure 4A) belonging to Figure 1A, the slope magnitude map (Figure 5B) of the $N_R(I_T,S_M)$ surface (Figure 4B) belonging to Figure 1B and the slope magnitude map (Figure 5C) of the $N_R(I_T,S_M)$ surface (Figure 4C) belonging to Figure 1C show triangular areas (blue) with low slope having their base parallel to the $I_T$ axis at $S_M \approx 20$. The triangular low-slope areas corresponding to the plateaus of the respective $N_R(I_T,S_M)$ surfaces taper towards large $S_M$ values. Moreover, the low-slope triangles have asymptotic tails with somewhat higher slope (red) protruding from the tips of the triangles away from the $I_T$ axis at $S_M \approx 80\text{-}100$ towards larger $S_M$ values and smaller $I_T$ values. These tails are the counterparts of ridges protruding from the plateaus of the $N_R(I_T,S_M)$ surfaces in Figure 4A-C.

The slope magnitude map of the $N_R(I_T,S_M)$ surface of Figure 1D shown in Figure 4D (Figure 5D) reveals that the flat foot-like structure is characterized by grooves and ridges running parallel to the $S_M$ axis, which are represented by stripes with a low magnitude of slope. These features mirror the corrugated $N_R(I_T)$ contour lines at fixed $S_M$ values in the foot region of the $N_R(I_T,S_M)$ surface shown in Figure 4D. This topography may be rationalized as follows. If $I_T$ is increased from zero to 255, at first very large contiguous entities appear that split up into a few still large sub-entities and various small objects extending a few pixels. While the small objects disappear as $I_T$ is further increased, the remaining larger sub-entities again split up in a few smaller yet relatively large objects and various small objects extending a few pixels. Interestingly, this behavior is invariant with respect to changes in $S_M$. Taking into ac-count the overall small $N_R$ values of the flat foot, the grooves and ridges oriented roughly parallel to the $S_M$ axis represent the properties of only a few imaged clusters of condensed nanorods.

**Determination of $N_{IF}$ values for weakly to intermediately condensed nanorod arrays**

The question arises as to how $N_{IF}$ values representing the imaged number of array elements can be extracted from the $N_R(I_T,S_M)$ surfaces displayed in Figure 4. On condition that imaged nanorod arrays have, at least to some extent, retained a signature of their initial topography, as it is the case for the nanorod arrays seen in Figure 1A-C, corresponding micrographs will display subpopulations of objects with similar properties. Size and brightness of objects belonging to such a subpopulation, for example tips of separate nanorods, make them distinguishable from random background noise or from other types of imaged objects. If so, these distinct subpopulations of imaged objects must dominate the image properties within specific $I_T$ and $S_M$ ranges. In the corresponding regions of the $N_R(I_T,S_M)$ surfaces, the number of recognized discrete objects varies, if at all, only slightly. The low-slope triangles in Figure 5A-C are such regions representing separated nanorods. Area and flatness of these plateaus decrease from Figure 5A to Figure 5B to Figure 5C. The areas of the thin low-slope protrusions emanating from the high-$S_M$ tips of the plateaus increases from Figure 5A to Figure 5B to Figure 5C along with the fraction of bunched nanorods discernible in Figure 1A, Figure 1B and Figure 1C and can be ascribed to small nanorod aggregates.

The low-slope triangles apparent in the $N_R(I_T,S_M)$ surfaces (Figure 4A-C) belonging to the SEM images of Figure 1A-C can be defined as stationary states that represent the number of imaged array elements $N_{IF}$. The question arises as to which of the $N_R(I_T,S_M)$ values within the low-slope triangles represents $N_{IF}$ in the best way and how this $N_R(I_T,S_M)$ value can be identified unequivocally. We suggest that the local slope minima on the $N_R(I_T,S_M)$ surfaces with the highest $N_R$ values are the best representations of $N_{IF}$. As obvious from the slope magnitudes maps (Figure 5A-C) of the $N_R(I_T,S_M)$ surfaces displayed in Figure 4A-C, the $N_R(I_T,S_M)$ surfaces may exhibit several local slope minima. There are most likely two rea-sons for the occurrence of these local slope minima. First, as $I_T$ increases, larger agglomerates of condensed nanorods appearing as a single entity may dissolve into several smaller objects, yet having sizes still exceeding $S_M$. The resulting increase in $N_R$ may locally balances the decrease in $N_R$ related to objects merging into the background. This effect may largely explain the occurrence of the tail of the plateau of the $N_R(I_T,S_M)$ surface apparent in Figure 5C. Secondly, local slope minima occurring as $S_M$ is increased may be related to the frequency density of nanorod agglomerate sizes. Previous reports suggest that clusters of condensed nanorods may have sizes that preferentially lie within a certain size range (Kong et al., 2006). If so, nanorod clusters having such preferred sizes would form small subpopulations of objects with similar properties, which would in turn be the origin of local slope minima along the $S_M$ direction, as apparent in Figure 5A and B. However, only the local slope minima located closest to the bases of the low-slope triangles mark the positions that represent $N_{IF}$, and exactly these local slope minima are the slope minima with the highest $N_R$ values.

Figure 6 shows the frequency density $N_P$ of the sizes (number of contiguous white pixels) $S_P$ of discrete objects contained in Figure 1A-C at the $I_T$ values belonging to $N_{IF}$, $I_T(IF)$ for $S_M = 0$ pixels. Note that $S_P$ is the actual number of pixels an identified object consists of, whereas $S_M$ is the threshold value for the minimum number of contiguous pixels a recognized object must consist of. Naturally, $S_P$ is larger than or equal to $S_M$. A large number of objects consists of one white pixel or a few contiguous white pixels. With increasing object size, $N_P$ steeply decreases and reaches minima at $S_P$ values between 20 and 40. Then, $N_P$ steeply increases to reach a distinct maximum at $S_P$ values of 80-90 corresponding to the tip areas of separate nanorods. In Figure 6B and C, certain numbers of objects extending up to several hundred contiguous white pixels appear (out of the range in Figure 6) that represent condensed nanorods.

As $S_M$ is increased along the $N_R(I_T(IF),S_M)$ contour lines in the $N_R(I_T,S_M)$ surfaces belonging to the fix $I_T(IF)$ value, all objects with $S_P$ values smaller than $S_M$ are discarded, while the sum

$$\sum_{i=S_M}^{\infty} N_{P,i}$$

of all objects with $S_P$ values equal to or larger than $S_M$ corresponds to $N_R(I_T(IF),S_M)$. $S_M(IF)$, the $S_M$ value belonging to $N_{IF}$, approximately corresponds to the $S_P$ value at which the minimum between high $N_P$ values at small $S_P$ values representing pixel noise and the $N_P$ maximum representing single nanorod tips occurs in the $N_P(S_P)$ profiles. Hence, if $S_M$ on the $N_R(I_T,S_M)$ surfaces is varied around $S_M(IF)$, $N_R$ changes only slightly. However, $N_R$ steeply decreases if $S_M$ passes the $N_P(S_P)$ maximum representing separate nanorod tips.

**Determination of $N_{IF}$ values for strongly condensed nanorod arrays**

The determination of $N_{IF}$ values of strongly agglomerated nanorod arrays, such as that shown in Figure 1D, suffers from the lack of extended subpopulations of objects with comparable properties. Apparently, Figure 1D does not contain extended subsets of objects with similar properties. In the $N_R(I_T,S_M)$ surface of Figure 1D displayed in Figure 4D the characteristic, distinct low-slope triangle apparent in Figure 4A-C is, therefore, absent. However, as apparent from the corresponding slope magnitude map (Figure 5D), the $N_R(I_T,S_M)$ surfaces derived from Figure 1D show several local slope minima, the occurrence of which likely results from the same effects as the occurrence of local slope minima in the slope magnitude maps derived from Figure 1A-C (Figure 5A-C). For Figure 1D $N_{IF}$ can now still be determined by application of the same criteria as for Figures 1A-C. Thus, $N_{IF}$ corresponds to the largest $N_R$ value along the sharp kink in the $N_R(I_T,S_M)$ surface between the $S_M$ range dominated by pixel noise and flat foot-like protrusion. The point on the $N_R(I_T,S_M)$ surface belonging to $N_R(I_T,S_M) = N_{IF}$ must then be the local slope minimum with the largest $N_R$ value and can be considered as a collapsed triangular low-slope region. It should also be noted that highly ordered arrays with a small area density of array elements may exhibit well-controlled agglomeration of nanorods yielding nanorod clusters rather uniform in size (Pokroy et al., 2009b; Chandra et al., 2008). These clusters would constitute a separate subpopulation of objects represented by a low-slope plateau in the $N_R(I_T,S_M)$ surfaces appearing at higher $S_M$ values than a plateaus originating from separated nanorods. Nevertheless, it is straightforward to assume that also in this case the $N_R$ value at the local slope minimum with the highest $N_R$ value is the best estimate of the number of imaged objects.

**Conclusions**

We suggest the integrity fraction *IF* as a quantitative measure of the structural integrity of nanorod arrays. The integrity fraction *IF* is the number of array elements $N_{IF}$ apparent in a micrograph of a nanorod array divided by the number $N_0$ of array elements a defect-free array would contain in the same image field. The difference $N_{IF} - N_0$ can thus be considered as a measure of the fraction of condensed nanorods. The integrity fraction *IF* is accessible by real-space analysis of microscopic images, which can be automated. As example in case, we analyzed a series of SEM images of polymer nanorod arrays differing only in the length of the nanorods. At first, a surface representing the number $N_R$ of identifiable objects in an image is generated as a function of $I_T$ and $S_M$, where $I_T$ is the minimum pixel intensity and $S_M$ the minimum pixel size an object must have to be counted. The $N_R(I_T,S_M)$ surfaces show a sharp kink roughly parallel to the $I_T$ axis that separates a steep region at small $S_M$ values dominated by pixel noise from a region at higher $S_M$ values dominated by the properties of the imaged objects. Separate nanorods uniform in size may constitute subpopulations of imaged objects with similar brightness and pixel size that are clearly distinguishable from the image background and from other types of objects. Then, the $N_R(I_T,S_M)$ surfaces show triangular low-slope plateaus within which $N_R$ only slightly changes when $I_T$ and/or $S_M$ are changed. In the second step, the number $N_{IF}$ of imaged objects including tips of single nanorods and larger nanorod clusters is determined by identifying the highest $N_R$ value belonging to a local slope minimum. The local slope minima with the highest $N_R$ values are typically located at close to the sharp kink. If a $N_R(I_T,S_M)$ surface possesses a triangular low-slope plateau, the local slope minimum with the largest $N_R$ value will lie close to the sharp kink at a position within in the triangular low-slope plateau. $N_0$ can easily be calculated if lattice geometry and lattice constant of the ideal counterpart of an imaged array are known. The integrity fraction *IF* may allow quantitative comparisons of the structur-al integrities of different nanorod arrays as well as quantitative comparisons of the structural integrities of a specific nanorod array at different stages of its use. Moreover, we anticipate that different states of smart nanorod arrays transformable into each other by reversible switching (Sidorenko et al., 2007) can be characterized in this way. While we focused on SEM images, the methodology presented here can also be applied to images obtained by other microscopic techniques, such as optical microscopy and scanning force microscopy. Finally, we anticipate that the methodology reported here can also be transferred to other problems, for example, the quantification of the porosities of porous materials.

**Materials and methods**

*Sample preparation*

PMMA ($M_w$=936 kg mol$^{-1}$, $M_n$ = 889 kg mol$^{-1}$, PDI = 1.05) was purchased from Polymer Standards Service (Mainz, Germany). Self-ordered AAO with a pore diameter of 180 nm, a lattice constant of 500 nm and uniform pore depths ranging from 1 µm to 5 µm was prepared following the two-step anodization process reported by Masuda and co-workers (Masuda et al., 1998). The PMMA was placed on the surface of AAO heated to 220 °C and kept at this temperature for 24 h under vacuum and under application of a load of 160 g/cm$^2$. The samples were then cooled to room temperature at the natural cooling rate of the furnace. Thus, PMMA nanorods completely filling the volume of the AAO pores connected to a PMMA film on top of the AAO were obtained. The AAO was then etched using a solution of 1.8 g $CrO_3$ in 100 ml 6% $H_3PO_4$.

*SEM characterization*

Prior to SEM investigations, the samples were coated with a thin gold layer. SEM images were taken with a JEOL JSM 6510 microscope at accelerating voltages ranging from 3 kV to 5 kV using a secondary electron detector. The pixel size of the imaged areas was 1280 x 870 pixels. The image fields of panels A, B and C of Figure 1 displaying PMMA nanorod arrays with nanorod lengths of 1 µm, 1.5 µm and 2 µm correspond to an area of 25.6 x 17.4 µm²; the image field of Figure 1D displaying the PMMA nanorod array with a nanorod length of 5 µm corresponds to an area of 51.2 x 34.8 µm².

*Image analysis*

The SEM images of Figure 1 were analyzed using the program ImageJ 1.44p (http://imagej.nih.gov). Pixels sharing either common edges or common tips were considered contiguous. $N_R$ profiles were obtained for $S_M = k \cdot 20$ with $k = 0,1,2,3,...$ by determining $N_R$ at the given $S_M$ value for all $I_T$ values $0 \leq I_T \leq 255$. The $N_R(I_T,S_M)$ surfaces were smoothened by Gaussian filtering to prevent the subsequently calculated slope magnitude maps from being dominated by the local roughness of the $N_R(I_T,S_M)$ surfaces, which represents local rather than global properties of the analyzed images. Gaussian filtering using the program MATLAB (R2008b, The MathWorks) was carried out by convolution with a filter kernel whose elements were set according to

$$h(x,y) = \frac{1}{2\pi\sigma^2} \cdot \exp\left(-\frac{x^2+y^2}{2\sigma^2}\right),$$

where $x$ and $y$ denote the distance from the center of the filter matrix. The variance σ was set to 3.5 in all calculations.

Magnitude maps of the difference quotients of the $N_R(I_T,S_M)$ surfaces representing the magnitude of the slope were obtained using the program MATLAB (R2008b, The MathWorks) as follows:

$$|\Delta N_R| = \sqrt{\Delta A^2 + \Delta B^2}$$
$$\Delta A = \frac{1}{2}[N_R(I_T - 1, S_M) - N_R(I_T + 1, S_M)]$$
$$\Delta B = \frac{1}{2}[N_R(I_T, S_M - 1) - N_R(I_T, S_M + 1)]$$

All plots were prepared using the program Origin 6.0 Professional.

**Acknowledgement**

The authors thank H. Tobergte for technical support and the European Research Council (ERC-CoG-2014; project 646742 INCANA) for funding.

**Figures and Tables**

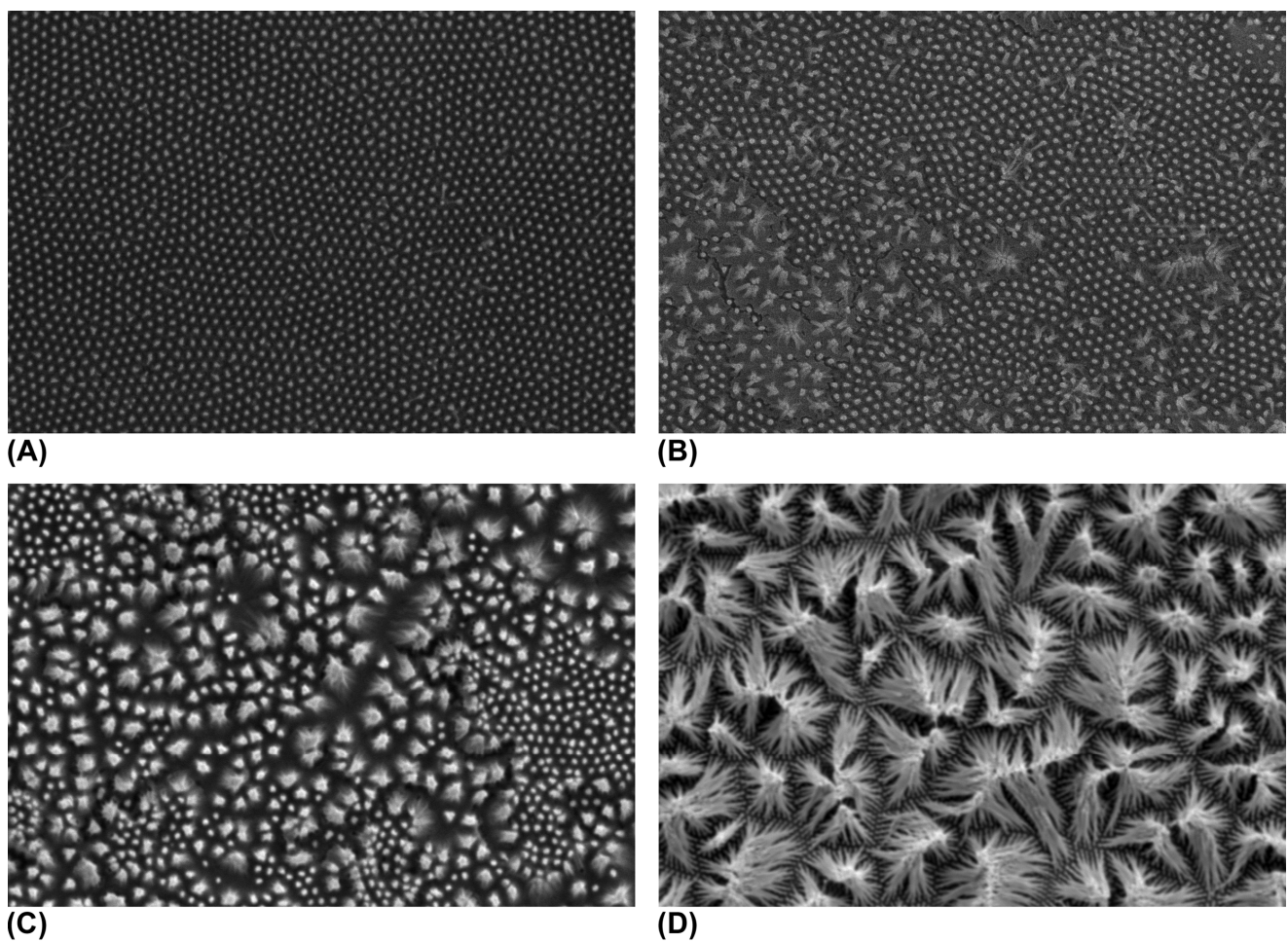

**Fig. 1.** Top-view SEM images composed of 1280 x 870 pixels showing PMMA nanorod arrays differing only in the length of the PMMA nanorods. The lengths of the nanorods amount to (A) 1 µm, (B) 1.5 µm, (C) 2 µm and (D) 5 µm. For panels (A), (B) and (C) the area of the image field is 25.6 x 17.4 µm², for panel (D) 51.2 x 34.8 µm².

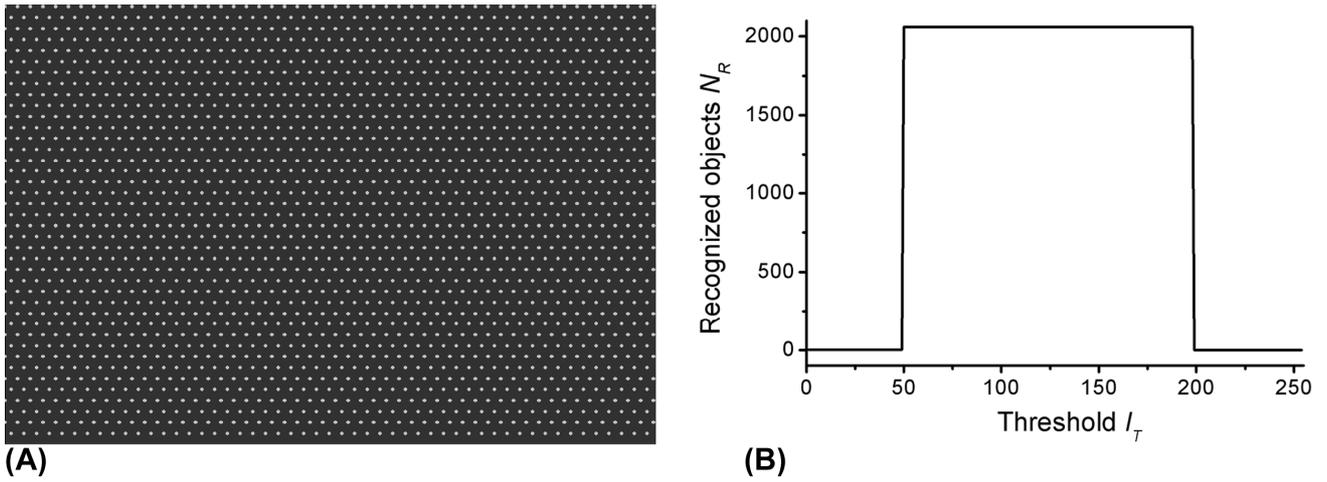

**Fig. 2.** (A) Schematic diagram showing an ideal hexagonal array. The pixel intensity of the background is 49, that of the array elements 198. (B) Corresponding plot of the number of recognized objects $N_R$ versus brightness threshold $I_T$. The steps occur at $I_T$ values of 49 and 198.

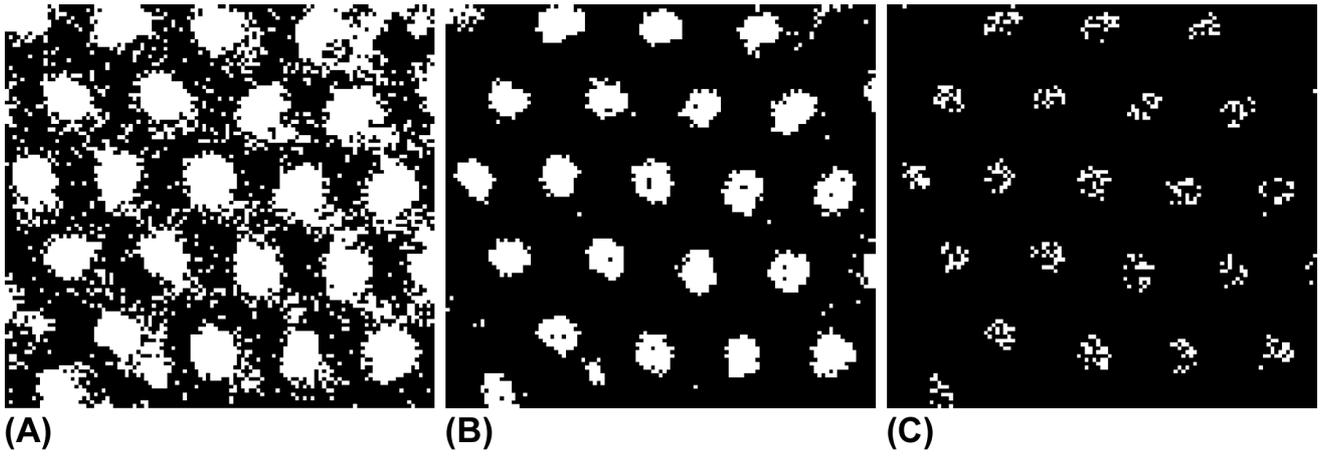

**Fig. 3.** Detail of Figure 1B showing an array of PMMA nanorods with a length of 1.5 μm for different brightness threshold values $I_T$. (A) $I_T = 67$; (B) $I_T = 95$; (C) $I_T = 133$.

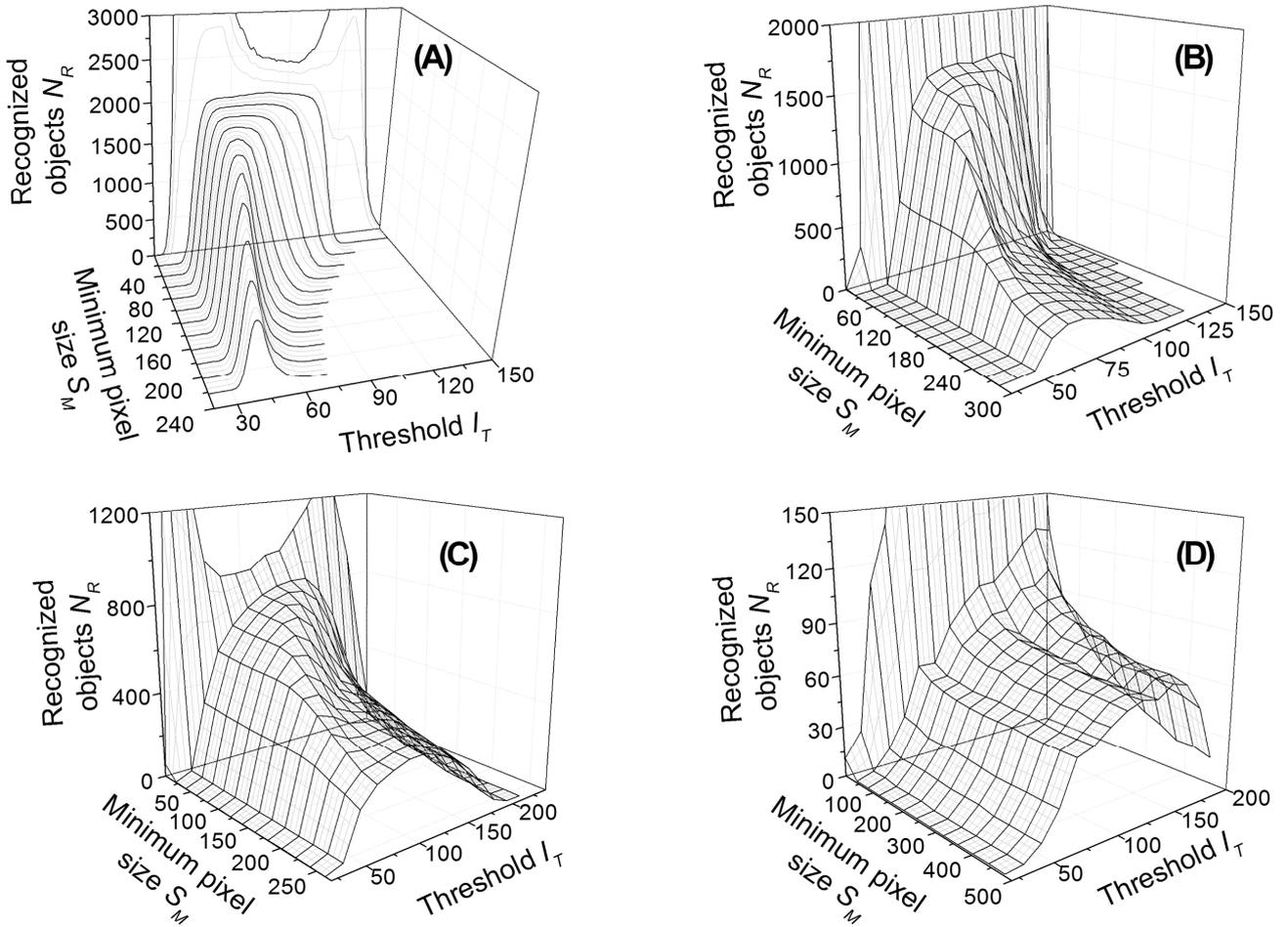

**Fig. 4.** Plots of the number of recognized objects $N_R$ as a function of brightness threshold $I_T$ and minimum object size $S_M$ obtained from the SEM images of Figure 1. (A) $N_R(I_T,S_M)$ surface belonging to Figure 1A; (B) $N_R(I_T,S_M)$ surface belonging to Figure 1B; (C) $N(I_T,S_M)$ surface belonging to Figure 1C; (D) $N(I_T,S_M)$ surface belonging to Figure 1D.

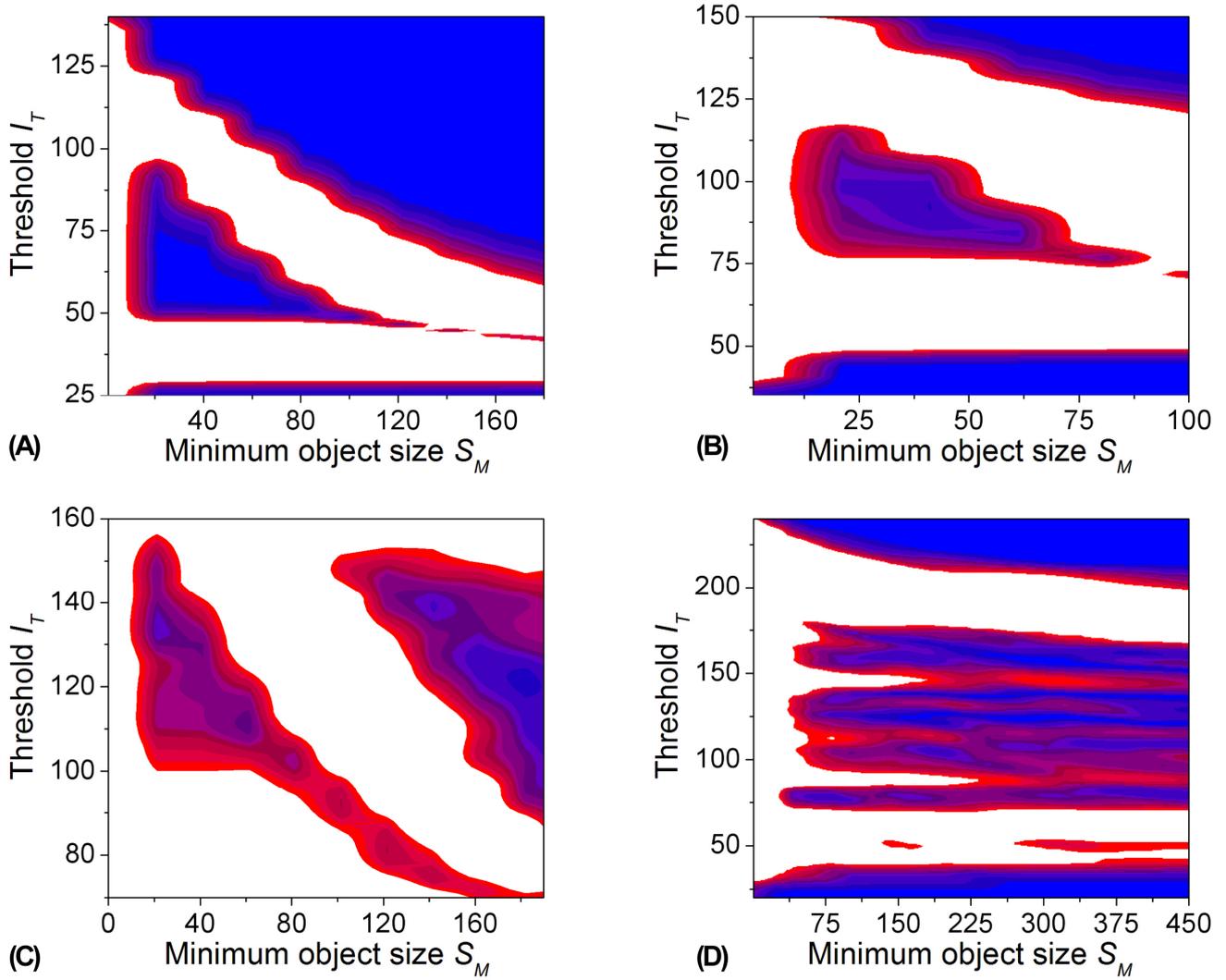

**Fig. 5.** Maps displaying the magnitude of the slope of the $N_R(I_T,S_M)$ surfaces displayed in Figure 4. The magnitude of the slope decreases from white to red to blue, i.e., blue denotes areas with the smallest magnitude of the slope. (A) Slope magnitude map of Figure 4A (representing Figure 1A); (B) slope magnitude map of Figure 4B (representing Figure 1B); C) slope magnitude map of Figure 4C (representing Figure 1C); (D) slope magnitude map of Figure 4D (representing Figure 1D). The step-like features of the level lines running diagonal with respect to the axes are artifacts related to the $S_M$ increment of 20 between the support lines of the $N_R(I_T,S_M)$ surfaces normal to the direction of the $S_M$ axis.

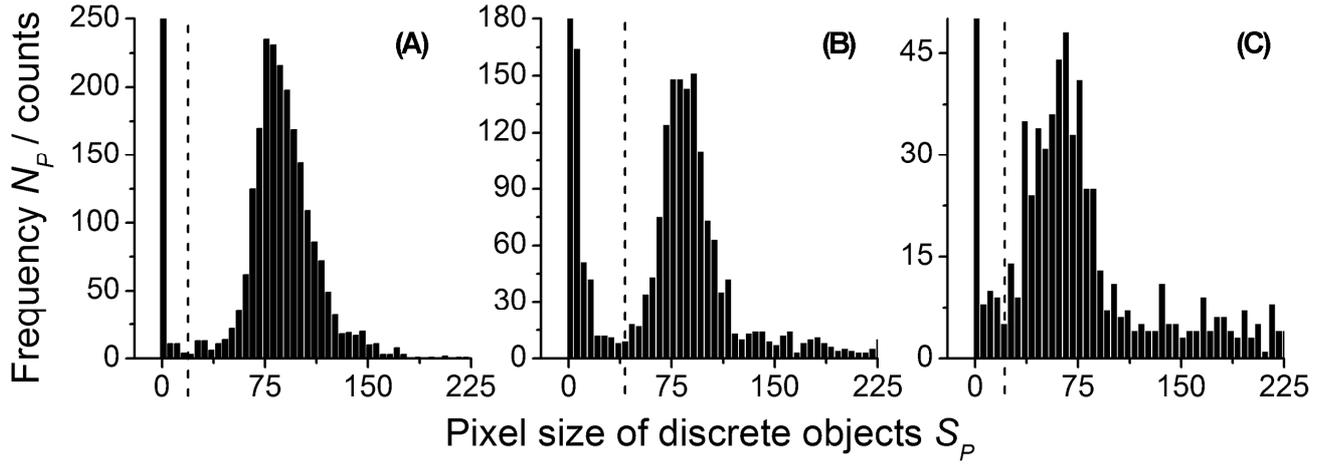

**Fig. 6.** Frequency density $N_P$ of the pixel sizes (number of contiguous white pixels) $S_P$ of discrete objects for $S_M = 0$ pixels at $I_T(IF)$, the $I_T$ value belonging to $N_{IF}$. Panel (A) was obtained from Figure 1A, panel (B) from Figure 1B and panel (C) from Figure 1C. The number of small objects consisting of one or only a few contiguous pixels exceeds the $N_P$ range represented by the $N_P$ axes. The dashed lines indicate $S_M(IF)$, the $S_M$ value belonging to $N_{IF}$.

| Figure | $L$ | $N_{IF}$ | $N_0$ | $IF$ |
|--------|-----|----------|-------|------|
| 1A | 1 µm | 2151 | 2056 | 1.04 |
| 1B | 1.5 µm | 1606 | 2056 | 0.78 |
| 1C | 2 µm | 864 | 2056 | 0.41 |
| 1D | 5 µm | 90 | 8228 | 0.02 |

**Table 1.** Results of real space analyses of the SEM micrographs displayed in Figure 1. *L*, length of nanorods; $N_{IF}$, array elements identified in the SEM images seen in Figure 1; $N_0$, number of array elements a defect-free array would contain in the same image field; *IF*, integrity fraction.

**References**

1. Beck, G. & Bretzler, R. (2011) Regularity of nanopores in anodic alumina formed on orientated aluminium single-crystals. *Mater. Chem. Phys.* **128**, 383-387.

2. Bernards, D. A. & Desai, T. A. (2010) Nanotemplating of Biodegradable Polymer Membranes. for Constant-Rate Drug Delivery. *Adv. Mater.* **22**, 2358–2362.

3. Boesel, L. F., Greiner, C., Arzt, E. & del Campo, A. (2010) Gecko-Inspired Surfaces: A Path to Strong and Reversible Dry Adhesives *Adv. Mater.* **22**, 2125-2137.

4. Chandra, D., Taylor, J. A., Yang, S. (2008) Replica molding of high-aspect-ratio (sub-)micron hydrogel pillar arrays and their stability in air and solvents. *Soft Matter* **4**, 979-984.

5. Chandra, D., Yang, S., Soshinsky, A. A. & Gambogi R. J. (2009) Biomimetic Ultrathin Whitening by Capillary-Force-Induced Random Clustering of Hydrogel Micropillar Arrays. *ACS Appl. Interf. Mater.* **1**, 1698-1704.

6. Chandra, D. & Yang, S. (2009) Capillary-Force-Induced Clustering of Micropillar Arrays: Is It Caused by Isolated Capillary Bridges or by the Lateral Capillary Meniscus Interaction Force? *Langmuir* **25**, 10430–10434.

7. Chandra, D. & Yang, S. (2010) Stability of High-Aspect-Ratio Micropillar Arrays against Adhesive and Capillary Forces. *Acc. Chem. Res.* **43**, 1080-1091.

8. Choi, M. K., Yoon, H., Lee, K. & Shin, K. (2011) Simple Fabrication of Asymmetric High-Aspect-Ratio Polymer Nanopillars by Reusable AAO Templates. *Langmuir* **27**, 2132–2137.

9. Del Campo, A. & Arzt, E. (2008) Fabrication Approaches for Generating Complex Micro- and Nanopatterns on Polymeric Surfaces. *Chem. Rev.* **108**, 911–945.

10. Feng, X. J. & Jiang, L. (2006) Design and Creation of Superwetting/Antiwetting Surfaces. *Adv. Mater.* **18**, 3063-3078.

11. Gitsas, A., Yameen, B., Lazzara, T. D., Steinhart, M., Duran, H. & Knoll W. (2010) Polycyanurate Nanorod Arrays for Optical-Waveguide-Based Biosensing. *Nano Lett.* **10**, 2173-2177.

12. Glassmaker, N. J., Jagota, A., Hui C.-Y. & Kim, J. (2004) Design of biomimetic fibrillar interfaces: 1. Making contact. *J. R. Soc. Interface* **1**, 23–33.

13. Grimm, S., Giesa, R., Sklarek, K., Langner, A., Gösele, U., Schmidt, H.-W., & Steinhart M. (2008) Nondestructive Replication of Self-Ordered Nanoporous Alumina Membranes via Cross-Linked Polyacrylate Nanofiber Arrays. *Nano Lett.* **8**, 1954-1959.

14. Grimm, S., Martin, J., Rodriguez, G., Fernandez-Gutierrez, M., Mathwig, K., Wehrspohn, R. B., Gösele, U., San Roman, J., Mijangos, C. & Steinhart, M. (2010) Cellular interactions of biodegradable nanorod arrays prepared by nondestructive extraction from nanoporous alumina. *J. Mater. Chem.* **20**, 3171-3177.

15. Haberkorn, N., Gutmann, J. S. & Theato, P. (2009) Template-Assisted Fabrication of Free-Standing Nanorod Arrays of a Hole-Conducting Cross-Linked Triphenylamine Derivative: Toward Ordered Bulk-Heterojunction Solar Cells. *ACS Nano* **3**, 1415-1422.

16. Hillebrand, R., Müller, F., Schwirn, K., Lee, W. & Steinhart, M. (2008) Quantitative Analysis of the Grain Morphology in Self-Assembled Hexagonal Lattices. *ACS Nano* **2**, 913–920.

17. Hillebrand, R., Grimm, S., Giesa, R., Schmidt, H.-W., Mathwig, K., Gösele, U. & Steinhart, M. (2009) Characterization of microrod arrays by image analysis. *Appl. Phys. Lett.* **94**, 164103.

18. Ho, A. Y. Y., Yeo, L. P., Lam, Y. C. & Rodríguez, I. (2011) Fabrication and Analysis of Gecko-Inspired Hierarchical Polymer Nanosetae. *ACS Nano* **5**, 1897–1906.

19. Jeong, H. E., Lee, S. H., Kim, P. & Suh K. Y. (2006) Stretched Polymer Nanohairs by Nanodrawing. *Nano Lett.* **6**, 1508-1513.

20. Jin, M. H., Feng, X. J., Feng, L., Sun, T. L., Zhai, J., Li, T. J. & Jiang, L. (2005) Superhydrophobic Aligned Polystyrene Nanotube Films with High Adhesive Force. *Adv. Mater.* **17**, 1977-1981.

21. Johnston-Peck, A. C., Wang, J. & Tracy, J. B. (2011) Formation and Grain Analysis of Spin-Cast Magnetic Nanoparticle Monolayers. *Langmuir* **27**, 5040–5046.

22. Kim, D.-H., Kim, P., Song, I., Cha, J. M., Lee, S. H., Kim B. & Suh, K. Y. (2006) Guided Three-Dimensional Growth of Functional Cardiomyocytes on Polyethylene Glycol Nanostructures. *Langmuir* **22**, 5419-5426.

23. Kim, T.-I., Jeong, H. E., Suh, K. Y. & Lee, H. H. (2009) Stooped Nanohairs: Geometry-Controllable, Unidirectional, Reversible, and Robust Gecko-like Dry Adhesive. *Adv. Mater.* **21**, 2276–2281.

24. Kong, J., Yung, K-L., Xu, Y., He, L., Lau, K. H. & Chan, C. Y. (2008) Self-Organized Micropatterns of High Aspect Ratio Polymer Nanofibers by Wetting of Nanopores. *J. Polym. Sci. Part B Polym. Phys.* **46**, 1280-1289.

25. Kwak, M. K., Pang, C., Jeong, H.-E., Kim, H.-N., Yoon, H., Jung, H.-S. & Suh, K.-Y. (2011) Towards the Next Level of Bioinspired Dry Adhesives: New Designs and Applications. *Adv. Funct. Mater.* **21**, 3606–3616.

26. Lee, H., Lee, B. P. & Messersmith, P. B. (2007) A reversible wet/dry adhesive inspired by mussels and geckos. *Nature* **448**, 338–341.

27. Mahdavi, A. et al. (2008) A biodegradable and biocompatible gecko-inspired tissue adhesive. *Proc. Natl. Acad. Sci. U. S. A.* **105**, 2307–2312.

28. Martin, C. (1994) Nanomaterials: A Membrane-Based Synthetic Approach. *Science* **266**, 1961–1966.

29. Masuda, H., Yada, K. & Osaka, A. (1998) Self-Ordering of Cell Configuration of Anodic Porous Alumina with Large-Size Pores in Phosphoric Acid Solution. *Jpn. J. Appl. Phys. Part 2 Lett.* **37**, L1340–L1342.

30. Mátéfi-Tempfli, S., Mátéfi-Tempfli, M. & Piraux, L. (2008) Characterization of nanopores ordering in anodic alumina. *Thin Solid Films* **516**, 3735-3740.

31. Matsunaga, M., Aizenberg, M. & Aizenberg, J. (2011) Controlling the Stability and Reversibility of Micropillar Assembly by Surface Chemistry. *J. Am. Chem. Soc.* **133**, 5545–5553.

32. Nielsch, K., Choi, J., Schwirn, K., Wehrspohn, R. B. & Gösele, U. (2002) Self-ordering regimes of porous alumina: The 10% porosity rule. *Nano Lett.* **2**, 677-680.

33. Pokroy, B., Epstein, A. K., Persson-Gulda, M. C. M. & Aizenberg, J. (2009a) Fabrication of Bioinspired Actuated Nanostructures with Arbitrary Geometry and Stiffness. *Adv. Mater.* 21, 463-469.

34. Pokroy, B., Kang, S. H., Mahadevan, L. & Aizenberg, J. (2009b) Self-Organization of a Mesoscale Bristle into Ordered, Hierarchical Helical Assemblies. *Science* **323**, 237-240.

35. Roca-Cusachs, P., Rico, F., Martínez, E., Toset, J., Farré, R. & Navajas D. (2005) Stability of Microfabricated High Aspect Ratio Structures in Poly(dimethylsiloxane). *Langmuir* **21**, 5542-5548.

36. Saez, A., Ghibaudo, M.; Buguin, A., Silberzan P. & Ladoux, B. (2007) Rigidity-driven growth and migration of epithelial cells on microstructured anisotropic substrates. *Proc. Natl. Acad. Sci. U. S. A.* **104**, 8281–8286.

37. Seiler, H. (1983) Secondary electron emission in the scanning electron microscope. *J. Appl. Phys.* **54**, R1-R18.

38. Sidorenko, A., Krupenkin, T., Taylor, A., Fratzl, P. & Aizenberg, J. (2007) Reversible Switching of Hydrogel-Actuated Nanostructures into Complex Micropatterns. *Science* **315**, 487-490.

39. Steinberg, T., Schulz, S., Spatz, J. P., Grabe, N., Mussig, E., Kohl, A., Komposch G. & Tomakidi, P. (2007) Early Keratinocyte Differentiation on Micropillar Interfaces. *Nano Lett.* **7**, 287-294.

40. Tanaka, T., Morigami, M. & Atoda, N. (1993) Mechanism of Resist Pattern Collapse during Development Process. *Jpn. J. Appl. Phys. Part 1* **32**, 6059–6064.

41. Tao, S. L. & Desai, T. A. (2007) Aligned Arrays of Biodegradable Poly($\varepsilon$-caprolactone) Nanowires and Nanofibers by Template Synthesis. *Nano Lett.* **7**, 1463-1468.

42. Vlad, A., Melinte, S., Mátéfi-Tempfli, M., Piraux, L. & Mátéfi–Tempfli, S. (2010) Vertical Nanowire Architectures: Statistical Processing of Porous Templates Towards Discrete Nanochannel Integration. *Small* **6**, 1974–1980.

43. Xue, L., Kovalev, A., Dening, K., Eichler-Volf, A., Eickmeier, H., Haase, M., Enke, D., Steinhart, M. & Gorb, S. (2013) Reversible adhesion switching of porous fibrillar adhesive pads by humidity. *Nano Lett.* **13**, 5541-5548.

44. Xue, L., Kovalev, A., Eichler-Volf, A., Steinhart, M. & Gorb S. (2015) Humidity-enhanced wet adhesion on insect-inspired fibrillar adhesive pads. Nat. Commun. 6, 6621.

45. Yanagishita, T., Nishio, K. & Masuda, H. (2005) Fabrication of metal nanhole arrays with high aspect ratios using two-step replication of anodic porous alumina. *Adv. Mater.* **17**, 2241-2243.

46. Zhang, Y., Lo, C.-W., Taylor, J. A. & Yang, S. (2006) Replica Molding of High-Aspect-Ratio Polymeric Nanopillar Arrays with High Fidelity. *Langmuir* 22, 8595-8601.